\documentclass{PoS}

\title{3D Relativistic MHD Simulations of Magnetized Spine-Sheath Relativistic Jets}
 
\ShortTitle{RMHD Simulations of Magnetized Spine-Sheath Relativistic Jets}

\author{\speaker{Y. Mizuno}$^{a}$, P. Hardee$^{b}$, and K.-I. Nishikawa$^{a}$\\
\llap{$^a$} National Space Science and Technology Center\\
             320 Sparkman Drive, VP 62, Huntsville, AL 35805, USA\\
E-mail: \email{Yosuke.Mizuno@msfc.nasa.gov}\\
\llap{$^b$} Department of Physics and Astronomy, The University of Alabama\\
             Tuscaloosa, AL 35487, USA\\
}

\abstract{We have performed numerical simulations of weakly and
strongly magnetized relativistic jets embedded in a weakly and
strongly magnetized stationary or mildly relativistic ($0.5c$) sheath
using the RAISHIN code. In the numerical simulations a jet with
Lorentz factor $\gamma=2.5$ is precessed to break the initial
equilibrium configuration. Results of the numerical simulations are
compared to theoretical predictions from a normal mode analysis of the
linearized RMHD equations describing a uniform axially magnetized
cylindrical relativistic jet embedded in a uniform axially magnetized
moving sheath. The prediction of increased stability of a
weakly-magnetized system with mildly relativistic sheath flow to
Kelvin-Helmholtz instabilities and the stabilization of a
strongly-magnetized system with mildly relativistic sheath flow is
confirmed by the numerical simulations.}

\FullConference{VI Microquasar Workshop: Microquasars and Beyond\\
		September 18-22 2006\\
		Societ\`a del Casino, Como, Italy}

\begin{document}

\section{Introduction}

Relativistic jets have been observed in galaxies and quasars
(AGNs)\cite{Urr95, Fer98}, in black hole binary star systems
(microquasars)\cite{Mir99}, and are thought responsible for the
gamma-ray bursts (GRBs)\cite{Mez06}. Proper motions observed in
microquasar and AGN jets imply the jet speeds from $\sim 0.9c$ up to
$\sim 0.999c$.

While jets at the larger scales may be kinetically dominated and
contain relatively weak magnetic fields, the possibility of much
stronger magnetic fields certainly exists closer to the acceleration
and collimation region. Recent GRMHD simulations of jet formation
performed by Mizuno et al. (2006a) \cite{Miz06a} indicate that highly
collimated high speed jets driven by the magnetic fields threading the
ergosphere may themselves reside with a broader wind or sheath outflow
driven by the magnetic fields anchored in the accretion disk. This
configuration might additionally be surrounded by a less collimated
accretion disk wind from the hot corona \cite{Nis05}.

Recent observations of high speed winds in several QSO's (0.1 - 0.4~c)
\cite{Pou04} also indicate that the highly relativistic jet could
reside in a high speed wind or sheath, at least close to the central
engine. Other observational evidence such as {\it limb brightening}
has been interpreted as evidence for a slower external flow
surrounding a faster jet spine \cite{Gio04}. A spine-sheath jet
structure has also been proposed based on theoretical arguments
\cite{Hen91}.

In this paper we investigate the stability properties of highly
relativistic jet flows allowing for the effects of strong magnetic
fields and relativistic flow in a sheath around the highly
relativistic jet by 3D RMHD numerical simulations. Such 3D numerical
simulations of fluid relativistic jets and accompanying theoretical
work has allowed relatively unambiguous interpretation of the
structures observed in the numerical simulations\cite{Har01, Agu01,
Har03}.

\section{Numerical Setup}

In order to study the long-term stability of magnetized sheath-spine
relativistic jets, we use the 3-dimensional GRMHD code ``RAISHIN''
with Cartesian coordinates in special relativity. The method is based
on a 3+1 formalism of the general relativistic conservation laws of
particle number and energy momentum, Maxwell equations, and Ohm's law
with no electrical resistance (ideal MHD condition) on a curved
spacetime \cite{Miz06b}. The RAISHIN code can perform special
relativistic computations in Minkowski spacetime by changing the
metric.

We consider the following initial conditions for the simulations:
``preexisting'' jet flow is established across the computational
domain. This represents the case in which a leading Mach disk and bow
shock has passed. The jet flow is surrounded by a low-density external
wind (medium). For all simulations, the ratio of densities is
$\rho_{j}/\rho_{e}=2.0$, where $\rho$ is the mass density in the
proper frame. The jet flow has $v_{j}=0.9165c$ and $\gamma \equiv (1 -
v^{2})^{-1/2}=2.5$. The initial magnetic field is assumed to be
uniform and parallel to the jet flow. We have performed two sets of
simulations. In the weakly magnetized (RHD) simulations, the relevant
sound speeds are $a_{e}=0.574c$ and $a_{j}=0.511c$, the relevant
Alfv\'{e}n speeds are $v_{Ae}=0.0682c$ and $v_{Aj}=0.064c$, and the
Alfv\'{e}n speed is much smaller than sound speed. In the strongly
magnetized (RMHD) simulations. The relevant sound speeds are
$a_{e}=0.30c$ and $a_{j}=0.226c$, the relevant Alfv\'{e}n speeds are
$v_{Ae}=0.56c$ and $v_{Aj}=0.45c$, and the Alfv\'{e}n speed is
comparable to the sound speed. In order to investigate the effect of
an external wind, we have performed simulations with no external wind
($v_{e}=0.0c$) and a mildly relativistic external wind ($v_{e}=0.5c$)
for both weakly- and strongly-magnetized cases. The computational
domain is $3 R_{j} \times 3 R_{j} \times 60 R_{j}$ with $60 \times 60
\times 600$ computational zones ( 10 computational zones span
$R_{j}$). We impose outflow boundary conditions on all surfaces except
the inflow plane, $z = 0$.  A precessional perturbation of angular
frequency $\omega R_{j}/v_{j}=0.93$ is applied at the inflow by
imposing a transverse component of velocity with
$v_{\bot}=0.01v_{j}$. The simulations are halted after $\sim 60$ light
crossing times of the jet radius,and before the perturbation has
crossed the entire computational region.

\section{Results}

\begin{figure}
\includegraphics[width=1.0\linewidth]{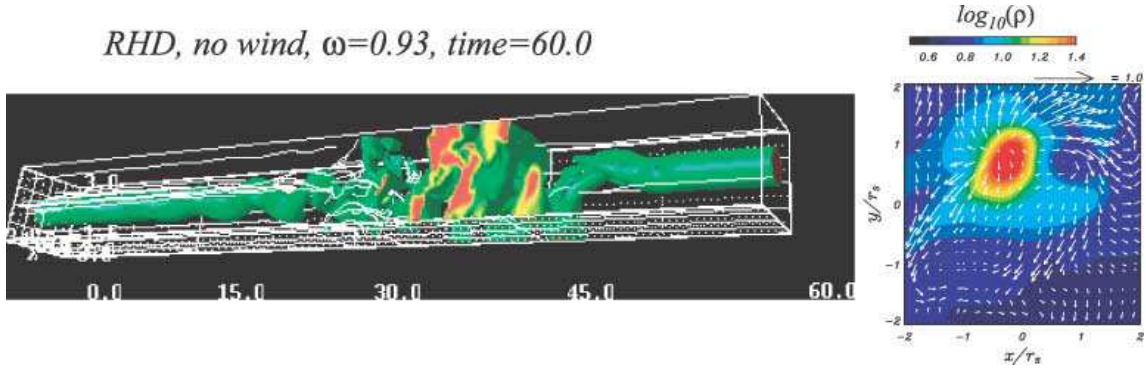}
\caption{Three-dimensional isovolume image ({\it a}) and transverse
({\it b}) cut ($z=30 R_{j}$) of the weakly magnetized case without
an external wind. The color scales show the logarithm of density. The
white lines indicate magnetic field lines. The arrows depict transverse
velocities. \label{f1}}

\end{figure}

Figure 1 shows the structure of a weakly magnetized jet
(three-dimensional isovolume image and transverse cuts) without an
external wind at $t=60$. The precession at the jet inflow plane
excites the Kelvin-Helmholtz (KH) instability. The initial
perturbation propagate down the jet and grows as a helical
structure. Beyond $R_{j}=40$ the KH instability disrupst the jet
structure strongly. The magnetic field is strongly distorted and
becomes complicated. The transverse cut at $z=30 R_{j}$ seen Fig 1b,
shows strong interaction with the external medium at this
distance. Transverse velocities show circular motions near the jet
surface caused by the helically twisted jet.

\begin{figure}
\centering
\includegraphics[width=1.0\linewidth]{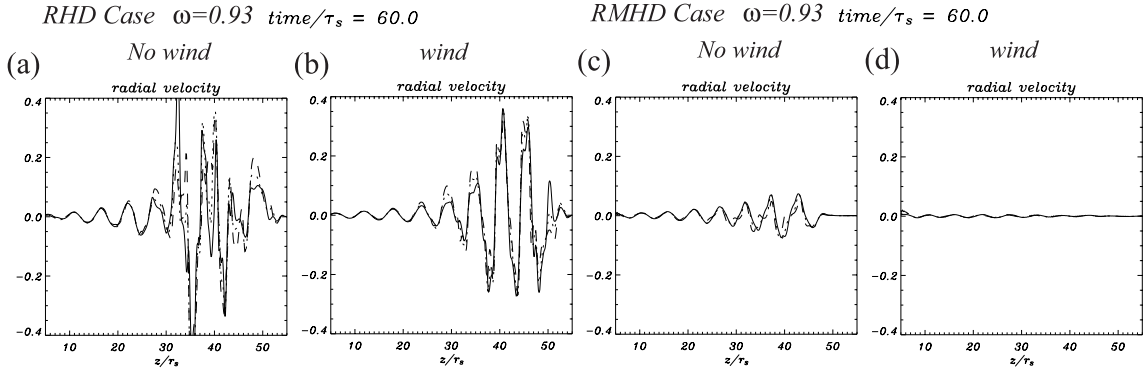}
\caption{Radial velocity ($v_{x}$) along the one-dimensional cuts
parallel to the jet axis located at $x/R_{j} =$ 0.2 (solid line), 0.5
(dotted line) and 0.8 (dashed line) for the weakly ({\it a, b}) and
strongly ({\it c, d}) magnetized cases without an external wind ({\it
a, c}) and with an external wind ({\it b, d}). \label{f2}}

\end{figure}

To investigate simulation results quantitatively, we take
one-dimensional cuts through the computational box parallel to the
z-axis at different radial distances along the transverse x-axis at
$x/R_{j}=$ 0.2, 0.5, and 0.8, as shown in Figure 2.

The weakly magnetized simulation results show velocity oscillation
from the growing helical KH instability. The simulations are fully
evolved to a non-linear phase by about $32 R_{j}$ for no external wind
and beyond about $38 R_{j}$ for an external wind. The external wind
has reduced the growth of KH instability and has delayed the onset of
the non-linear phase. The dominant wavelength of oscillation is
$\lambda/R_{j} \sim 6$ in both weakly magnetized cases. The strongly
magnetized simulation results without an external wind also show
growing velocity oscillation (Figure 2c), albeit more slowly. The
simulation is still evolving linearly and has not reached the
non-linear phase. From the comparison with the weakly-magnetized
cases, the stronger magnetic field reduces the growth rate of the
helical KH instability. The strongly magnetized simulation result with
an external wind reveals a damped oscillation. Therefore the external
wind in the strongly magnetized case leads to damping of KH
instability and stabilizes the jet. The dominant oscillation
wavelength is $\lambda/R_{j} \sim 5$ in both of the strongly
magnetized cases.

The simulations confirm theoretical predictions for the stability of a
jet spine-sheath or jet-wind configuration Hardee (2006) \cite{Har06}.
Theoretical predictions arise from a normal mode analysis of the
linearized ideal RMHD equations describing a uniform axially
magnetized cylindrical relativistic jet embedded in a uniform axially
magnetized moving sheath. The most important theoretical prediction
obtained from the normal mode analysis is that the presence of an
external wind and strong magnetic field can reduce the growth rate or
lead to damping of the KH instability. This prediction is confirmed by
the present simulation results.

\section{Conclusion}

We have performed numerical simulations of weakly and strongly
magnetized relativistic jets embedded in a weakly and strongly
magnetized stationary or mildly relativistic ($0.5c$) sheath using the
RAISHIN code \cite{Miz06b}. In the numerical simulations a jet with
Lorentz factor $\gamma=2.5$ is precessed to break the initial
equilibrium configuration. Results of the numerical simulations
have been compared to theoretical predictions from a normal mode analysis of
the linearized RMHD equations describing a uniform axially
magnetized cylindrical relativistic jet embedded in a uniform axially
magnetized moving sheath. The prediction of increased stability of
weakly-magnetized systems with mildly relativistic sheath flow to
Kelvin-Helmholtz instability and the stabilization of a
strongly-magnetized system with mildly relativistic sheath flow is
verified by the numerical simulation results.

\section{Acknowledgments}

Y. M. is a NASA Postdoctoral Program fellow at NASA Marshall Space
Flight Center. K. N. is partially supported by the National Science
Foundation awards ATM-0100997, INT-9981508, and AST-0506719, and the
National Aeronautic and Space Administration award
NASA-INTEG04-0000-0046 to the Univ of Alabama in Huntsville.
P.H. acknowledges partial support by National Space Science and
Technology (NSSTC/NASA) cooperative agreement NCC8-256 and NSF award
AST-0506666.  The simulations have been performed on the IBM p690 at
the National Center for Supercomputing Applications (NCSA) which is
supported by the NSF and Altix3700 BX2 at YITP in Kyoto University.


\begin{thebibliography}{}

\bibitem{Urr95} C.M.~Urry, \& P.~Padovani, \emph{Unified Schemes for
Radio-Loud Active Galactic Nuclei}, \emph{PASP} {\bf 107} (803) 1995.

\bibitem{Fer98} A.~Ferrari, \emph{Modeling Extragalactic Jets},
\emph{ARAA} {\bf 36} (539) 1998.

\bibitem{Mir99} I.F.~Mirabel, \& L.F.~Rodr\'{i}guez, \emph{Sources of
Relativistic Jets in the Galaxy}, \emph{ARAA} {\bf 37} (409) 1999.

\bibitem{Pir05} T.~Piran, \emph{The physics of gamma-ray bursts},
\emph{ Reviews of Modern Physics}, {\bf 76} (1143) 2005.

\bibitem{Mez06} P.~M\'{e}sz\'{a}ros, \emph{Gamma-ray bursts},
\emph{Rep. Prog. Phys.} {\bf 69} (2259) 2006.

\bibitem{Miz06a} Y.~Mizuno, K.-I.~Nishikawa, S.~Koide, P.~Hardee, \&
G.J.~Fishman, \emph{General Relativistic Magnetohydrodynamic
Simulations of Jet Formation with a Thin Keplerian Disk}, \emph{ApJL}
2006a submitted.

\bibitem{Nis05} K.-I.~Nishikawa, G.~Richardson, S.~Koide, K.~Shibata,
T.~Kudoh, P.~Hardee, \& G.J.~Fishman, \emph{A General Relativistic
Magnetohydrodynamic Simulation of Jet Formation}, \emph{ApJ} {\bf 625}
(60) 2005.

\bibitem{Pou04} K.~Pounds, \& K.~Page, \emph{Evidence for massive
ionised outflows in (super-Eddington?) AGN}, \emph{Nuc. Phys. B
Proc. Sup.} {\bf 132} (107) 2004.

\bibitem{Gio04} G.~Giovannini, \emph{Observational Properties of Jets
in Active Galactic Nuclei}, \emph{Ap\&SS} {\bf 293} (1) 2004.

\bibitem{Hen91} G.~Henri, \& G.~Pelletier, \emph{Relativistic
electron-positron beam formation in the framework of the two-flow
model for active galactic nuclei}, \emph{ApJL} {\bf 383} (L7) 1991.

\bibitem{Har01} P.E.~Hardee, P.A.~Hughes, A.~Rosen, \& E.~Gomez,
\emph{Relativistic Jet Response to Precession and Wave-Wave
Interactions}, \emph{ApJ} {\bf 555} (744) 2001.

\bibitem{Agu01} I.~Agudo, et al. \emph{Jet Stability and the
Generation of Superluminal and Stationary Components}, \emph{ApJL}
{\bf 549} (L183) 2001.

\bibitem{Har03} P.E.~Hardee, \& P.A.~Hughes, \emph{The Effect of
External Winds on Relativistic Jets}, \emph{ApJ} {\bf 583} (116) 2003.

\bibitem{Miz06b} Y.~Mizuno, K.-I.~Nishikawa, S.~Koide, P.~Hardee, \&
G.J.~Fishman, \emph{RAISHIN: A High-Resolution Three-Dimensional
General Relativistic Magnetohydrodynamics Code}, \emph{ApJS} 2006b
submitted.

\bibitem{Har06} P.E.~Hardee, \emph{The Stability Properties of
Strongly Magnetized Spine-Sheath Relativistic Jets}, 2006 in
preparation

\end{thebibliography}
\end{document}